\documentstyle[prb,aps]{revtex}
\begin{document}
\input psbox
\draft
\title{Exchange coupling between magnetic layers across non-magnetic
superlattices} 
\author{M. S. Ferreira}
\address{Department of Mathematics, Imperial College, London, 
SW7 2BZ, UK }
\date{\today}
\maketitle
\begin{abstract}
The oscillation periods of the interlayer exchange coupling are
investigated when two magnetic layers are separated by a metallic
superlattice of two distinct non-magnetic materials. In
spite of the conventional behaviour of the coupling as a function of the
spacer thickness, new periods arise when the coupling is looked upon as
a function of the number of cells of the superlattice. The new
periodicity results from the deformation of the corresponding Fermi
surface, which is explicitly related to a few controllable parameters,
allowing the oscillation periods to be tuned. 
\end{abstract}

\vspace{1cm}
\section{Introduction}

Oscillatory interlayer exchange coupling in metallic magnetic
multilayers causes the magnetizations of neighbouring magnetic layers
separated by non-magnetic spacers to be spontaneously aligned
ferromagnetically or antiferromagnetically depending on the thickness of
the spacer material. This phenomenon has been intensively studied
both experimentally and theoretically over the last few years
\cite{review1,review2}. 

The relation between the periods, by far the most
investigated feature of the oscillations, and the electronic structure
of multilayered systems is a fundamental aspect of the
mechanism responsible for the origin of this phenomenon. In fact, the 
periods were originally shown to be dependent on the shape of the spacer
Fermi surface, where two distinct criteria were proposed for
determining the periods with which the interlayer coupling oscillates
\cite{ed1,bruno1,bruno2}. 
Within the quantum well theory the wave vectors yielding the effective
oscillation periods correspond to extremal radii of the Fermi surface in
the direction perpendicular to the layers, or in other words to half the
caliper measurements \cite{ed1}. On the other hand, in the RKKY theory
adapted to the layered geometry, the periods are associated with
wave vectors perpendicular to the layers linking two points of the Fermi
surface with antiparallel velocities \cite{bruno1,bruno2}. Despite the
distinction between those geometrical criteria, the periods 
predicted by both theories, at least in their original formulations,
agree in most cases. When they do not coincide, such as, for example,
when the lattice lacks reflection symmetry about the layer planes, suitable
transformations of the Fermi surface can be performed, which
reestablish the correspondence between the quantum well and the RKKY
periods \cite{fcc111}. Recently however, fundamental 
oscillation periods not directly associated with the shape of the spacer
Fermi surface were obtained \cite{non-rkky,fundamental,fe/cu}. Those
periods, which correspond to higher order contributions to the exchange
coupling, are no longer determined by the above geometrical criteria but
by a set of more general conditions on the electronic structure of the
system, although they are still indirectly dependent on the shape of the
Fermi surface. These conditions provide a systematic way of determining
the oscillation periods involved in the interlayer coupling. 

In order to test the applicability of those general conditions, it is worth
investigating them in situations where the spacer Fermi surface, clearly a
relevant factor in this matter, is variable. One approach is to use an
alloy spacer of variable composition although the Fermi surface is no
longer sharply defined. Also, it is well know that the scattering of
electrons by random impurities strongly damps the oscillations. In fact, 
the amplitude of the coupling is found to decay exponentially with the
concentration of impurities \cite{paper1,okuno}. An alternative to an
alloy is a metallic superlattice whose Fermi surface depends on the
proportions and layer thicknesses of the constituent materials but where
the oscillations are not damped by incoherent electronic scattering. 
Moreover, nature provides us with only a limited number of non-magnetic
metals. However, by combining distinct materials in superlattices the
number of possible non-magnetic superlattices is infinite. 

The purpose of this communication is to investigate the oscillation
periods of the interlayer exchange coupling when the spacer is a
non-magnetic superlattice. It is shown that the alteration of the
oscillation periods reflects the deformation of the corresponding Fermi
surface only when the coupling is investigated as a function of the
number of superlattice cells separating the magnetic layers. In the
simple model treated here, the oscillation periods of the coupling are
written in terms of controllable parameters which are related to the
composition of the superlattices, and which, accordingly, allow the
periods to be tuned. It is hoped that such a simple model reflects what
occurs in realistic systems, and this may be confirmed by future
experiments. 

The sequence adopted in this work is described as follows.
In order to make the work self-contained, a brief summary of the general
method for determining the oscillation periods in a multi-orbital
tight-binding model \cite{fundamental} is presented in section
\ref{periods}. Secondly, the method is shown to be applicable when the
spacer is a superlattice, using the clear correspondence between orbital
and intracell indices. Subsequently, the oscillation periods are
obtained following the analysis of the electronic structure of the
system and are explicitly written in terms of parameters associated with
the superlattice composition.

\section{Oscillation periods}
\label{periods}

A brief summary of the general method for determining the
oscillation periods of the interlayer coupling is presented here. See 
reference \cite{fundamental} for a comprehensive discussion of the
method.

For the sake of simplicity, we consider two parallel magnetic planes,
labelled $0$ and $n$, embedded in an infinite non-magnetic metal.
As far as the coupling as a function of the spacer thickness is
concerned, the number of magnetic planes influences only the phase and
amplitude of the oscillations and does not affect the periods
\cite{zeetal}. Since the periods are the main concern here, the
restriction to two single magnetic planes does not pose any limitations
on the results obtained. 

The interlayer coupling ${\cal J}$ is given by the difference in total
energy between the antiferromagnetic and ferromagnetic configurations
or, in other words, by the energy necessary to flip the magnetization in
one of the planes. 
Based on the formalism introduced by d'Albuquerque e Castro {\it et al},
the coupling ${\cal J}$ is written in terms of the one-electron Green
functions of the multilayered system by the following expression
\begin{equation}
{\cal J} \,=\, \sum_{{\bf q}_\parallel} \,
\int d\,\omega\,{\cal F}({\bf q}_\parallel,\omega)\,\,\,,
\label{j}
\end{equation}
where ${\cal F}({\bf q}_\parallel,\omega) = f(\omega) \, Im \, Tr \, \ln
\left\{ 1 + 4 \, G_{n 0}^\uparrow({\bf q}_\parallel,\omega) \, V_{ex} \,
G_{0 n}^\downarrow({\bf q}_\parallel,\omega) \, V_{ex} \right\}$. 
In the equations above $G_{n 0}^\sigma({\bf q}_\parallel,\omega)$ is the
propagator between planes $0$ and $n$ for electrons with spin $\sigma$
in the ferromagnetic configuration of the system, $f(\omega)$ is the
Fermi function, $V_{ex}$ is a diagonal matrix in orbital indices
representing the strength of the local exchange potential in the
ferromagnetic layers, and the sum over ${\bf q}_\parallel$ is restricted
to the two-dimensional Brillouin zone. Bearing in mind that the exchange
splitting of the magnetic material is localized on the two planes $0$
and $n$, the propagators $G$ can be rewritten as a function of the bulk
spacer propagators $g_{n 0}$ and $g_{0 n}$, i.e., 
\begin{equation}
G_{n 0}^\sigma({\bf q}_\parallel,\omega) = G^\sigma \left(g_{n 0}({\bf
q}_\parallel,\omega) ,  g_{0 n}({\bf q}_\parallel,\omega) \right) \,\,. 
\label{Gg}
\end{equation}
Moreover, it can be shown that within the single-band model the
dependence of $G^\sigma$ upon those propagators is such that $G_{n
0}^\sigma = G^\sigma \left(g_{n 0} \, g_{0 n} \right)$. 
It is clear that the spacer thickness dependence of the
interlayer coupling is entirely determined by the off-diagonal
propagators $g_{n 0}$ and $g_{0 n}$. The investigation of the
oscillation periods thus requires an analysis of $g_{n 0}$ as a function
of the interplane distance. 

In the layered geometry, where ${\bf q}_\parallel$ represents the
in-plane wave-vector, the expression for a general matrix element
$g_{\ell m}^{\mu \nu}$ representing the propagation from an orbital
$\mu$ at plane $\ell$ into an orbital $\nu$ at plane $m$ of an electron
at the Fermi level $E_F$ is given by
\begin{equation}
g_{\ell m}^{\mu \nu}({\bf q}_\parallel,E_F) = ({d \over 2\pi}) \,
\sum_s \, \int_{-{\pi \over d}}^{\pi \over d}\,\,dq_\perp \, {a_{s
\mu}^*(q_\perp) \, a_{s \nu}(q_\perp) \, e^{-i q_\perp (\ell-m) d}
\over E_F - E_s({\bf q}_\parallel,q_\perp) + i\,0^+}\,\,,
\label{integral}
\end{equation}
where the integration variable $q_{\perp}$ is the wave vector
perpendicular to the layers, $d$ is the
interplane distance, $E_s({\bf q}_\parallel,q_\perp)$ describes the bulk
spacer band structure, $s$ is the band index, and $a_{s \mu}(q_\perp)
\equiv \langle q_\perp \, s \, \vert q_\perp \, \mu \rangle$ is the
projection of the eigenstate $s$ into the orbital $\mu$ for a given
$q_\perp$. By changing the integration contour from the real axis to the
boundaries of a semi-infinite rectangle in the complex plane whose base
lies on the real axis between $-\pi / d$ and $\pi / d$, the evaluation
of the integral above is simplified to the determination of the poles of
the integrand inside the appropriate contour and their respective
residues. The rectangle is in the upper-half plane for $\ell < m$ and in
the lower-half plane otherwise. Clearly, the poles are given by 
the values of $q_\perp$ satisfying the condition ${E_F}^+ - E_s({\bf 
q}_\parallel,q_\perp) = 0$, where ${E_F}^+ = E_F + i\,0^+$. 

By adding the residues associated with the contributory poles, labelled
$q_j^s$, an analytical expression for $g_{\ell m}$ is obtained, {\it
i.e.}, 
\begin{equation}
g_{\ell m}({\bf q}_\parallel,\omega) = \sum_s \, \sum_{j} A_{s j} \,
e^{-i q_j^s (\ell-m) d}\,\,\,,
\label{glm}
\end{equation}
where the matrix elements of the matrix $A_{s j}$ are
\begin{equation}
A_{s j}^{\mu \nu}({\bf q}_\parallel,\omega) =  \,-\,i\,d\,\,a_{s
\mu}^*(q_j^s)\,a_{s \nu}(q_j^s)\left\{\left[\,{\partial E_s({\bf
q}_\parallel,q_\perp) \over \partial q_\perp}\,\right]
_{q_\perp=q_j^s}\right\}^{-1}\,\,.
\label{AA} 
\end{equation}
The coefficients $A_{s j}^{\mu \nu}$ depend neither on $\ell$ nor on $m$
and the only dependence on the distance between the planes is inside the
argument of the exponentials. It should be stressed that this simple
method for obtaining the one-electron propagators is exclusive to the
one-dimensional case, and therefore for fixed values of ${\bf
q}_\parallel$. The exponentials in Eq.(\ref{glm}) are independent of the
orbital indices, which means that all matrix elements oscillate with the
same periods. Note that in calculating Eq.(\ref{AA}) it was assumed that
the coefficients $a_{s \mu}(q_\perp)$ have neither singularities nor
branch points inside the integration contour. In this way, the
coefficients are evaluated just at the poles $q_j^s$. In general, branch
cuts associated with the coefficients $a_{s \mu}(q_\perp)$ may exist, but
because they may be chosen not to intersect the real axis, the factor
$e^{-i q_\perp (\ell - m) d}$ in the integrand of Eq.(\ref{integral})
ensures that the corresponding contribution is negligible for large
$\vert \ell - m \vert$. Therefore, the expression above is correct in
the asymptotic limit of large interplane distance \cite{fundamental}. 

It is clear from Eq.(\ref{glm}) that $g_{\ell m}$ is an oscillatory
function of the distance between the planes and it oscillates with
different periods, each one associated with its respective wave vector
$q_j^s$. For this reason it is interesting to look at the physical
significance of the poles. The poles may be real or complex, but we note
that only the real poles contribute significantly to the coupling. This
is because complex poles, when inserted in Eq.(\ref{glm}), damp the
oscillations and are important only in cases of very thin spacers. Real
poles, which are obtained from the 
electronic structure of the bulk spacer, are just the
perpendicular coordinates of the Fermi surface for a fixed value of
${\bf q}_\parallel$. In other words, they are the perpendicular
components of the wave vectors with which electrons of energy $E_F$
propagate across the spacer.

Having investigated $g_{\ell m}$ as a function of the interplane
distance, it is worth stressing that the functions ${\cal F}$ must
oscillate with the same periods. When expanded in a Fourier series, 
taking into account the possibility of quasi-periodicity of $g_{n 0}$
\cite{non-rkky,fundamental,zeetal}, the function ${\cal F}$ is written
as 
\begin{equation}
{\cal F} = \sum_{(m_1^1,...,m_j^s,...)}\, {\cal
C}_{(m_1^1,...,m_j^s,...)} \,\, e^{i \, n  \, d \,
\sum_{s j} (m_j^s \, q_j^s)}\,\,, 
\label{f}
\end{equation}
where ${\cal C}_{(m_1^1,...,m_j^s,...)}$ are the Fourier coefficients
determining the amplitude of the oscillations and depend on the exact
form of the function ${\cal F}$. The indices $(m_1^1,...,m_j^s,...)$
are integers and there are as many as the number of poles $q_j^s$. Note
that the periods are given by $\lambda_{(m_1^1,...,m_j^s,...)} = 2 \pi /
(\sum_{s j} m_j^s q_j^s)$ and do not depend on the exact form of ${\cal
F}$, since they are obtained from the bulk spacer band 
structure. Because the indices $m_j^s$ run over the set of integers,
there is always an infinite number of periods defined by Eq.(\ref{f}),
even for a small number of poles $q_j^s$. In spite of the somewhat
congested notation of the general expression above, in practice there
are only a few indices involved due to the restricted number of real
solutions crossing the Fermi level. In fact, the number of solutions is
related to the number of sheets of the Fermi surface. 

When Eq.(\ref{f}) is inserted into Eq.(\ref{j}), the two-dimensional
integral can be evaluated through the stationary phase approximation,
valid in the asymptotic limit of large spacer thickness
\cite{ed1,newspa}. In doing so, the periods of the coupling are
identified by selecting the periods of ${\cal F}$ associated with wave
vectors satisfying the following condition, 
\begin{equation}
\sum_{s j} \left[\,m_j^s \; {\bf \nabla}_\parallel \, q_j^s({\bf
q}_\parallel,E_F)\right] \, = 0 \,\,,
\label{grad}
\end{equation}
where ${\bf \nabla}_\parallel$ is the two-dimensional gradient in ${\bf
q}_\parallel$-space. The equation above gives
the necessary conditions for constructive interference between
electrons across the spacer. The otherwise destructive interference does
not contribute to the coupling. Eq.(\ref{grad}) simply selects the
points across the two-dimensional Brillouin zone which yield the
effective periods with which the coupling oscillates. 

As pointed out in reference \cite{fundamental}, although the wave
vectors $q_j^s$ are directly related to the Fermi surface, the general
wave vectors $\sum_{s j} (m_j^s \, q_j^s)$ are not. This is a result of
the interference between incommensurate wave vectors $q_j^s$ giving rise
to a new set of periods not directly obtained from the structure of the
Fermi surface. Consequently, geometrical criteria cannot be used in
general to select the oscillation periods. However, for Fermi
surfaces with a single sheet, where the off-diagonal propagators
oscillate as a function of the interplane distance with only one
period, the selection rules given by Eq.(\ref{grad}) coincide with the
geometrical criteria of the RKKY theory \cite{bruno1,bruno2}. The only
difference in this case is that Eq.(\ref{grad}) determines the
respective harmonics, in addition to the fundamental RKKY periods. 

In summary, the determination of the oscillation periods results from
the analysis of the bulk spacer off-diagonal propagators where the
relevant real wave vectors, under the stationary phase condition, are
obtained from the band structure. When dealing with multi-sheet Fermi
surfaces, the possibility of interferences between incommensurate wave
vectors may introduce new features in the oscillations, which are not
present in the case of single-period oscillations of the propagators. 

\section{Superlatticed spacers}

\begin{figure}
\begin{center}\mbox{\psboxscaled{600}{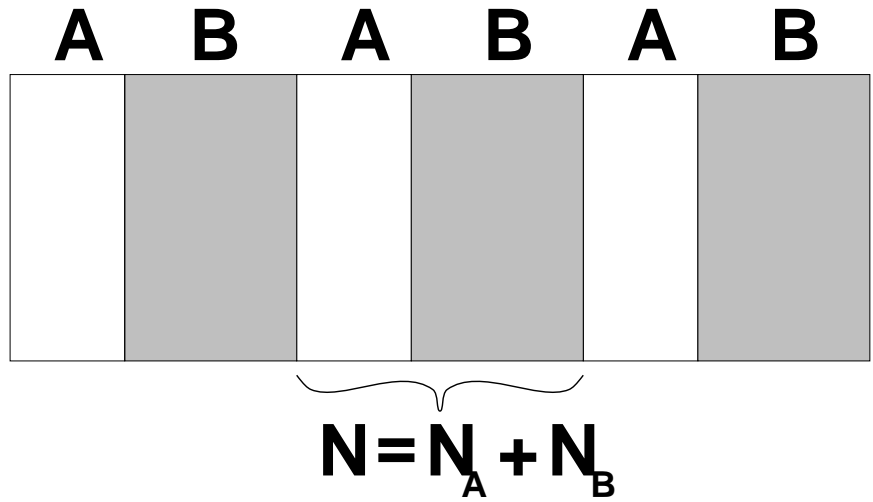}}\end{center}
\caption{Schematic representation of a superlatticed spacer.}
\label{superlattice}
\end{figure}

Having summarized the method for determining the oscillation periods of
the coupling, it is worth stressing that no particular form for the
spacer material was assumed. The only information required is the bulk
band structure which makes possible the investigation of the
coupling across arbitrary spacers. Spacers consisting of superlattices
of distinct non-magnetic materials will be investigated here. More
specifically, the material into which the magnetic planes are embedded
is composed of two non-magnetic metals $A$ and $B$ with respective
thicknesses $N_A$ and $N_B$ periodically alternated, as schematically
displayed in figure \ref{superlattice}. In order to make a distinction, 
these structures will be hereafter named superlatticed spacers, in
contrast to the conventional single-layered spacers. The interlayer
coupling across superlatticed spacers can be looked upon as a
function of different thicknesses, namely $N_A$, $N_B$, $N$ and $N_c$,
where $N_c$ represents the number of cells separating the magnetic
planes. Each cell is formed by a set of $N_A$ planes of metal $A$ and
$N_B$ planes of metal $B$. As far as the oscillation periods are
concerned, the 
coupling across superlatticed spacers as a function of $N_A$ or $N_B$
does not present new features when compared to the coupling across
conventional single-layered spacers as a function of $N_A$ or $N_B$,
respectively. In other words, if a single-layered spacer of material $A$
yields an interlayer coupling with period $\lambda_A$ and likewise
$\lambda_B$ is the period for the coupling across a single-layered
spacer of material $B$, the coupling across a superlatticed spacer of
materials $A$ and $B$ will oscillate with period $\lambda_A$ when seen
as a function of $N_A$ and with period $\lambda_B$ when seen as a
function of $N_B$. However, when looked upon as a function of $N_c$, for
fixed values of $N_A$ and $N_B$, the coupling oscillates with periods
different from $\lambda_A$ and $\lambda_B$. Although the superlattice
Fermi surface is different from the individual Fermi surfaces of
materials A and B separately, it is only when investigated as a function
of $N_c$ that the deformation of the Fermi surface is translated into a
change in the periods. 

Following the steps of the previous section, the bulk spacer band
structure must be calculated. For the sake of simplicity, the
single-band tight-binding model is considered assuming a simple cubic
lattice structure for the superlattice grown along the (100) direction. 
For the two given materials $A$
and $B$, the in-plane energies are labelled $\epsilon_A$ and $\epsilon_B$,
and the interplane hoppings are $t_A$ and $t_B$. The hopping between
planes $A$ and $B$ is represented by $t_{AB}$. We recall that the
tight-binding parameters above depend on the in-plane 
wave vector ${\bf q}_\parallel$ through $\epsilon_\beta({\bf
q}_\parallel) = \epsilon_\beta^0 + 2 t_\beta^0 \left[ \cos(q_x d) +
\cos(q_y d) \right]$ and $t_\beta({\bf q}_\parallel) = t_\beta^0$, where
$\beta$ is either $A$ or $B$, $\epsilon_\beta^0$ is the on-site energy
and $t_\beta^0$ is the hopping between nearest-neighbour sites in
material $\beta$. Note that $q_z$ is perpendicular to the layers whereas
$q_x$ and $q_y$ are the coordinates of ${\bf q}_\parallel$.

Because each set of $N$ planes is periodically repeated, where
$N = N_A + N_B$, the system can be represented by cells of $N$
planes equally spaced. Consequently, the tight-binding Hamiltonian of
the infinite multilayered system, when written in the appropriate basis,
becomes a $N \times N$ matrix given by  
\begin{equation}
{\cal H} = \left( \begin{array}{c c}
\left[{\cal H}_A(N_A)\right] & \left[{\cal H}_{AB}\right] \\
\left[{\cal H}_{BA}\right] & \left[{\cal H}_B(N_B)\right] \\
\end{array} \right) \,\,,
\label{H(NxN)}
\end{equation}
where $\left[{\cal H}_A(N_A)\right]$ and $\left[{\cal H}_B(N_B)\right]$
are square matrices of order $N_A$ and $N_B$, respectively.
The sub-matrices $\left[{\cal H}_A(N_A)\right]$ and
$\left[{\cal H}_B(N_B)\right]$ are explicitly given by
\begin{equation}
\left[{\cal H}_\beta(N_\beta)\right] = \left. \left(
\begin{array}{c c c c} 
\epsilon_\beta\; & t_\beta \; & 0\; & 0\; \\
t_\beta\; & \epsilon_\beta\; & t_\beta\; & 0\; \\
0\; & t_\beta\; & \epsilon_\beta\; & t_\beta\; \\
0\; & 0\; & t_\beta\; & \epsilon_\beta\; \\
\end{array} \right) \,\,\, \right\} \,\, {N_\beta} \,\,\,\,.
\end{equation}
Note that $\left[{\cal H}_\beta(N_\beta)\right]$ is a tri-diagonal 
matrix equivalent to the single-band tight-binding Hamiltonian of a 
slab consisting of $N_\beta$ planes. The sub-matrices off the diagonal 
in Eq.(\ref{H(NxN)}) are $N_A \times N_B$ and $N_B \times N_A$ matrices 
such that
\begin{equation}
\left[{\cal H}_{AB}\right] = t_{AB} \underbrace{ \left(
\begin{array}{c c c c}
0\; & 0 \; & 0\; & e^{- i \,q_\perp \, N \, d} \\
0\; & 0 \; & 0\; & 0\; \\
1\; & 0\; & 0\; & 0\; \\
\end{array} \right) }_{N_B} \left. \begin{array}{c} \\ \\ \\
\end{array} \right\} N_A
\label{Hab}
\end{equation}
and $\left[{\cal H}_{BA}\right] = \left[\left({\cal
H}_{AB}\right)^\dagger\right]$. In the matrix directly above $q_\perp$
is the wave vector perpendicular to the layers and $d$ is the interplane
distance, which is assumed to be the same for both materials. The basis
producing such matrices results from a Fourier transform 
of the localized-atomic-orbital basis along the direction perpendicular
to the layers. Since the cells of $N$ planes are equally spaced by $N
d$, the associated wave vector $q_\perp$ is in the range
$[{-\pi \over N d} , {\pi \over N d}]$, that is, the corresponding
Brillouin zone along this direction is reduced by a factor $N$. The
tight-binding Hamiltonian above is equivalent to the
$N$-band Hamiltonian of a one-dimensional system. In other words, the
$N$ intracell planes may be considered as orbitals of a linear chain
whose nearest neighbour elements are $N d$ apart. Having established
this correspondence, hereafter intracell and orbital indices will be 
mentioned with no distinction between them.

Following section \ref{periods}, the oscillation periods result from
the analysis of the bulk spacer off-diagonal propagators as a function
of the distance between the planes. Note however that in the case
treated here the effective interplane distance must be the distance
between the cells, {\it i.e.}, $D = N d$. Hence, the off-diagonal
propagators as a function of the number of cells $N_c$ are the ones to
be investigated. These propagators depend on the band structure of the
system, as clearly shown in Eq.(\ref{integral}). The original
tight-binding Hamiltonian is partially diagonalized by both in-plane and
perpendicular Fourier transforms, the remaining problem being the
determination of the eigenvalues of the $N \times N$ matrix in
Eq.(\ref{H(NxN)}). Therefore, the information on the oscillation periods
of the coupling in a superlatticed system depends on the
diagonalization of this matrix, which will be presented in the next
section. 

To maintain the correspondence between the intracell
planes of the superlattice and orbital indices of the linear chain, one
has to assume that in the case of superlatticed spacers the magnetic
planes embedded in the non-magnetic system must be formed by two
separate magnetic cells. However, for the sake of simplicity and with no
loss of generality, we consider only two magnetic planes. If
we label the magnetic planes with the intracell indices $\alpha$ in the
cell $0$ and $\alpha^\prime$ in the cell $N_c$, the relevant Green
functions to be calculated are the off-diagonal propagators $g_{0
N_c}^{\alpha \alpha^\prime}({\bf q}_\parallel,\omega)$, i.e.,
\begin{equation}
g_{0 N_c}^{\alpha \alpha^\prime}({\bf q}_\parallel,\omega) = ({D
\over 2\pi}) \, \sum_s \, \int_{-{\pi \over D}}^{\pi \over D}\,\,dq_\perp
\, {a_{s \alpha}^*(q_\perp) \, a_{s \alpha^\prime}(q_\perp) \, e^{ i
q_\perp N_c D} \over \omega - E_s({\bf q}_\parallel,q_\perp) +
i\,0^+} \,\,.
\label{off}
\end{equation}
In addition to the fact that $\alpha$ and $\alpha^\prime$ are
intracell indices, the only difference between the equation above and
Eq.(\ref{integral}) is the interplane distance $D$ which modifies the
integral limits. Similarly to section \ref{periods}, the integral
above is evaluated through the residues method, where the determination
of the poles follow the diagonalization of the Hamiltonian. 

\subsection{Electronic structure}

Making use of the particular symmetry of the matrices displayed
above, the determinant of ${\cal H}$ in Eq.(\ref{H(NxN)}) can be written
in terms of the sizes of its sub-matrices $N_A$ and $N_B$ yielding a
simple analytical expression for the oscillation periods of the
one-electron propagators. The energy eigenvalues $\omega$ are solutions
of the equation $\det |{\cal H} - \omega \hat{{\bf 1}} | = 0$, where
$\hat{{\bf 1}}$ is the identity matrix. In this case the sub-matrices
along the diagonal become    
\begin{equation}
\left[{\cal H}_\beta(N_\beta) - \omega \hat{{\bf 1}} \right] =
\underbrace{ 
\left( \begin{array}{c c c c} 
\epsilon_\beta - \omega\; & t_\beta \; & 0\; & 0\; \\
t_\beta\; & \epsilon_\beta - \omega\; & t_\beta\; & 0\; \\
0\; & t_\beta\; & \epsilon_\beta - \omega\; & t_\beta\; \\
0\; & 0\; & t_\beta\; & \epsilon_\beta - \omega\; \\
\end{array} \right) }_{N_\beta} \,\,\,\,,
\end{equation}
whereas the block matrices off the diagonal are still given by
$\left[{\cal H}_{AB}\right]$ and $\left[{\cal H}_{BA}\right]$. 

The determinant of a $N \times N$ square matrix is a sum of $N!$
terms. Those terms are made of products of $N$ matrix elements,
following all their possible permutations. Because the sub-matrices on
the diagonal have only two distinct non-zero elements, the 
evaluation of their determinants by counting the number of combinations
allowed by the possible permutations is not particularly difficult and
can be expressed in term of the sizes $N_\beta$. The solution of this
combinatorial problem is given by \cite{combinatorial}
\begin{equation}
\det | {\cal H}_\beta(N_\beta) - \omega \hat{{\bf 1}} | =
\sum_{j=0}^{N_\beta/2} \left(-1\right)^j 
\left(\begin{array}{c}
N_\beta - j \\ j 
\end{array}\right) \;(\epsilon_\beta - \omega)^{N_\beta - 2j}
\;(t_\beta)^{2j} \,\,,
\label{det-tri}
\end{equation}
where $j$ is an integer, the upper limit of the sum is the integer part
of $N_\beta/2$ and 
\begin{equation}
\left(\begin{array}{c}
N_\beta - j \\ j 
\end{array}\right) = {( N_\beta - j ) ! \over ( N_\beta - 2 j ) ! \,\, j
!}
\end{equation}
is the binomial coefficient. Note that the expression above is valid for
symmetric tri-diagonal matrices with similar elements along each of the
diagonals. Bearing in mind that ${\cal H}_\beta(N_\beta)$ is equivalent
to the slab Hamiltonian, the values of $\omega$ satisfying $\det | {\cal
H}_\beta(N_\beta) - \omega \hat{{\bf 1}} | = 0$ through
Eq.(\ref{det-tri}) are just the associated eigenvalues.

The Hamiltonian ${\cal H}$ in Eq.(\ref{H(NxN)}) is almost tri-diagonal. 
The matrix ${\cal H}$ is not tri-diagonal due to two of its elements,
namely ${\cal H}_{1,N}$ and ${\cal H}_{N,1}$, which are
non-zero. Therefore, an expression for the determinant of ${\cal H}$ is
not expected to be very different from the one in Eq.(\ref{det-tri}). In
fact, by considering the only two non-zero elements of the submatrices
in Eq.(\ref{Hab}) and recounting the number of possible permutations,
the determinant of the matrix ${\cal H}$ can be expressed in terms of
the determinants of the matrices ${\cal H}_A(N_A)$ and ${\cal
H}_B(N_B)$. In the similar case of ${\cal H} - \omega \hat{{\bf 1}}$,
the determinant is given by 
\begin{eqnarray}
\det | {\cal H} - \omega \hat{{\bf 1}} | & = & \det | {\cal H}_A(N_A) -
\omega \hat{{\bf 1}} | \times \det | {\cal H}_B(N_B) - \omega \hat{{\bf
1}} | \nonumber \\
        &  & - \; 2 \left(t_{AB}\right)^2 \, \det | {\cal H}_A(N_A-1) -
\omega \hat{{\bf 1}} | \times \det | {\cal H}_B(N_B-1) - \omega
\hat{{\bf 1}} | \nonumber \\
                       &   &  + \; \left(t_{AB}\right)^4 \, \det | {\cal
H}_A(N_A-2) - \omega \hat{{\bf 1}} | \times \det | {\cal H}_B(N_B-2) -
\omega \hat{{\bf 1}} | \nonumber \\
        &  & + \; \left(t_{AB}\right)^2
\left(t_A\right)^{N_A-1}\left(t_B\right)^{N_B-1} \cos(q_\perp N \, d)
\,\,. 
\label{detH}
\end{eqnarray}
Note that the equation above is written in terms of lower-order
determinants of Eq.(\ref{det-tri}) and is an analytical expression for 
$\det | {\cal H} - \omega \hat{{\bf 1}} |$ in terms of its sizes $N_A$
and $N_B$. 

It is instructive to point out that when $A = B$, that is,
when a single-layered spacer is considered, the usual eigenvalues
$\omega(q_\perp) = \epsilon + 2 t \cos(q_\perp d)$ of an infinite linear
chain are obtained from Eq.(\ref{detH}) through the condition $\det |
{\cal H} - \omega \hat{{\bf 1}} | = 0$.

The eigenvalues $\omega$ of the Hamiltonian ${\cal H}$ are solutions of 
the polynomial equation defined by Eqs.(\ref{detH}) and (\ref{det-tri})
through the condition $\det |{\cal H} - \omega \hat{{\bf 1}} | = 0$.
To illustrate this point, figure \ref{bandn3n6} displays the band
structure of a superlatticed system consisting of two materials $A$ and
$B$ whose tight-binding parameters are arbitrarily chosen as
$\epsilon_A^0 = \epsilon_B^0 = 0$, $t_A = -1/2$ and $t_B = -1/3$. Two
cases are shown in figure \ref{bandn3n6}, namely $N = 3$ and $N = 6$. In
each case a number of possible values for $N_A$ and $N_B$ is presented.
Due to the equivalence between the single-band-cellular and
multi-band linear Hamiltonians, it is clear in all the cases that, 
for a given $N_A$, there are $N$ solutions per $q_\perp$ forming the
respective bands. All bands are separated by energy gaps except for the
ones in figure \ref{bandn3n6}A corresponding to $N_A = 0$ and $N_A =
3$. Those two cases correspond to single materials $B$ and $A$,
respectively.  

Since we are interested in determining the oscillation periods
of the one-electron propagators, what is needed from the polynomial
equation defined by Eqs.(\ref{detH}) and (\ref{det-tri}) is the wave
vector $q_\perp$ with which electrons with energy $\omega$ propagate
across the superlatticed spacer, that is, the poles $q_\perp$ of the
integrand in Eq.(\ref{off}). Whereas the determination of the
eigenvalues requires finding the zeros of an $N$-th degree polynomial in
$\omega$, the calculation of the wave vector $q_\perp$ for a given
energy $\omega$ is straightforward due to the simple dependence of
Eq.(\ref{detH}) on $q_\perp$. The advantage of writing the expression of
the determinant in terms of the number of planes $N_A$ and $N_B$ becomes
clear, for the oscillation periods can be written in terms of those
parameters. Furthermore, bearing in mind that the wave vectors $q_\perp$
are the perpendicular components of the spacer Fermi surface, we can
describe how the Fermi surface is deformed with the composition of the
cells. The wave vector $q_\perp$ is such that
\begin{eqnarray}
\cos(q_\perp N \, d) & = & {-1 \over
\left(t_A\right)^{N_A}\left(t_B\right)^{N_B}} \left\{
\det | {\cal H}_A(N_A) -
\omega \hat{{\bf 1}} | \times \det | {\cal H}_B(N_B) - \omega \hat{{\bf
1}} | \right. \nonumber \\
        &  & - \; 2 \, t_A \, t_B \, \det | {\cal H}_A(N_A-1) -
\omega \hat{{\bf 1}} | \times \det | {\cal H}_B(N_B-1) - \omega
\hat{{\bf 1}} | \nonumber \\
  &   &  \left. + \; t_A^2 \,\, t_B^2  \, \det | {\cal H}_A(N_A-2) -
\omega \hat{{\bf 1}} | \times \det | {\cal H}_B(N_B-2) - \omega
\hat{{\bf 1}} | \right\} \,\,,
\label{cosq}
\end{eqnarray}
where, for simplicity, the interplane hopping $t_{AB}$ was assumed to be
the geometrical average between $t_A$ and $t_B$. Note that
for a given energy $\omega$ (and ${\bf q}_\parallel$) there is only one
value of $\cos(q_\perp N \, d)$ defined. This indicates that the
off-diagonal propagators oscillate as a function of the number of cells
$N_c$ with a single period. In this case the Fermi surface is
deformed as a function of $N_A$ and $N_B$ but the number of periods is
left unchanged. This can be understood through the size reduction of the
Brillouin zone along the perpendicular direction $q_z$, folding the
Fermi surface back into the zone when it crosses the boundaries in $q_z
= \pm {\pi \over N d}$. 

\subsection{Deformation of the Fermi surface}

\begin{figure}
\begin{center}\mbox{\psboxscaled{700}{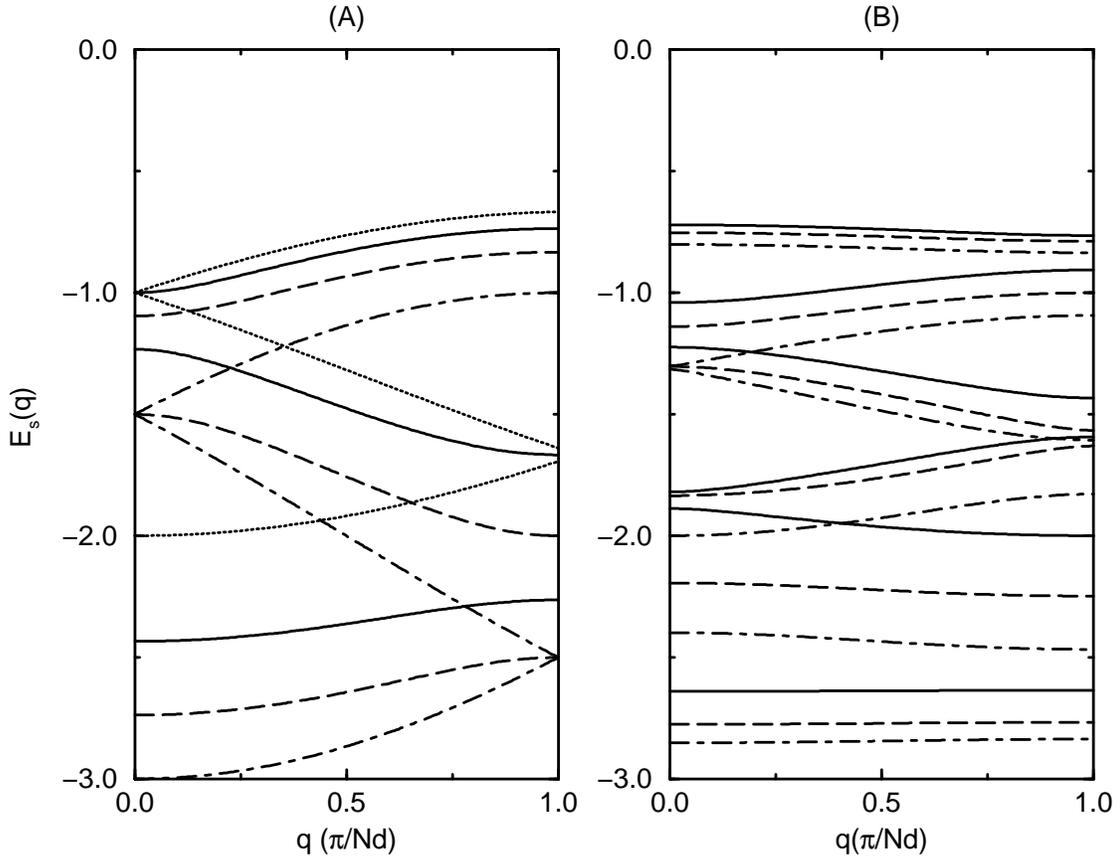}}\end{center}
\caption{Band structures $E_s(q_\perp, q_x = q_y = 0)$ for (A) $N =
3$ and (B) $N = 6$ as a function of $q_\perp$. In figure A the dotted line
is $N_A = 0$, the full line is $N_A = 1$, the dashed line is $N_A = 2$
and the dot-dashed line is $N_A = 3$. In figure B the the full line is
$N_A = 2$, the dashed line is $N_A = 3$ and the dot-dashed line is $N_A
= 4$. }
\label{bandn3n6}
\end{figure}

\begin{figure}
\begin{center}\mbox{\psboxscaled{1200}{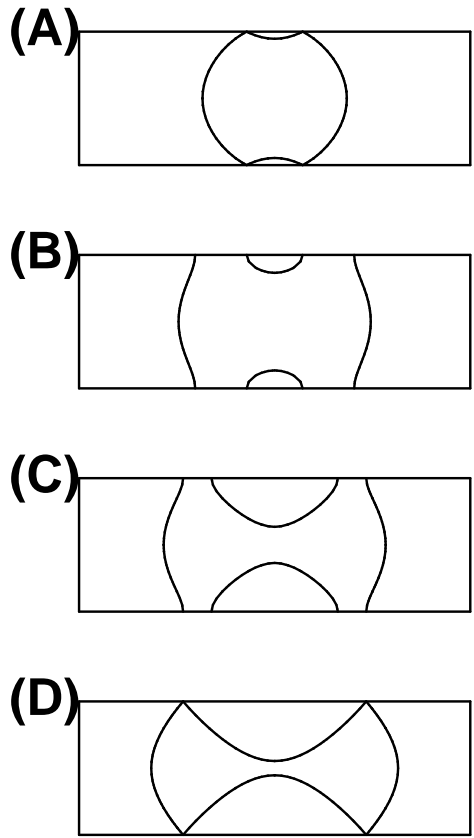}}\end{center}
\caption{Fermi surfaces for $N = 3$. (A) $N_A = 0$, (B) $N_A = 1$,
(C) $N_A = 2$ and (D) $N_A = 3$. Note that figures (A) and (D) correspond
to the bulk Fermi surface of the single-material spacers $B$ and $A$,
respectively. The rectangles are the Brillouin zones, whose dimensions
are $2 \pi/d$ horizontally and $2 \pi/3 d$ vertically.}
\label{FSN3}
\end{figure}

\begin{figure}
\begin{center}\mbox{\psboxscaled{1200}{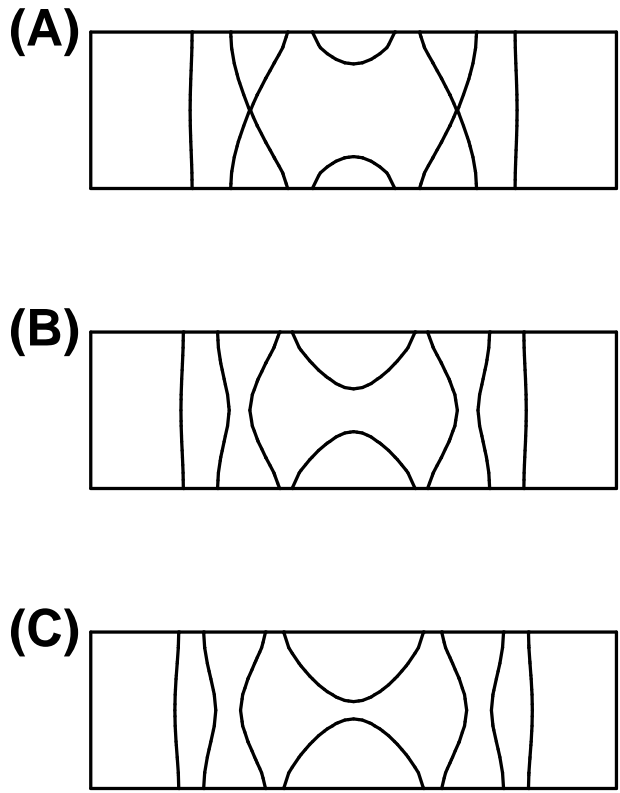}}\end{center}
\caption{Fermi surfaces for $N = 6$. (A) $N_A = 2$, (B) $N_A = 3$
and (C) $N_A = 4$. The rectangles are the Brillouin zones, whose
dimensions are $2 \pi/d$ horizontally and $2 \pi/6 d$ vertically.}
\label{FSN6}
\end{figure}

Here, given two materials $A$ and $B$, we shall look at the change 
of the Fermi surface for different thicknesses $N_A$ and $N_B$. 
Still assuming the tight-binding parameters used in the evaluation of
figure \ref{bandn3n6}, the cross sections of the superlatticed spacer Fermi
surface for $q_x = 0$ are shown in figures \ref{FSN3} and
\ref{FSN6}. Two cases are displayed, namely $N = 3$ with $E_F = -1.6$ in
figure \ref{FSN3} and $N = 6$ with $E_F = -1.35$ in figure \ref{FSN6},
where $E_F$ represents the Fermi level. In all those figures the
vertical axes are along the direction perpendicular to the layers
whereas the  horizontal direction represents $q_y$. The Fermi surfaces
are drawn inside the first Brillouin zone represented by the rectangles,
which are reduced along the $z$ direction by a factor $N$. Note that
figures \ref{FSN3}A ($N_A = 0$) and \ref{FSN3}D ($N_A = 3$) correspond
to the bulk Fermi surfaces of the pure materials $B$ and $A$,
respectively. The choice of different Fermi levels for $N = 3$ and $N =
6$ results from the analysis of figure \ref{bandn3n6}, using the
criterion of selecting a Fermi level which crosses the most number of
bands. In fact, in figure \ref{bandn3n6}A $E_F = -1.6$ yields solutions
for all possible values of $N_A$ and $N_B$ whereas in figure
\ref{bandn3n6}B $E_F = -1.35$ intersects the curves $N_A = 2$, $N_A = 3$
and $N_A = 4$. 

As discussed in section \ref{periods}, in the case where only one period
exists in the propagators, the RKKY geometrical picture coincides with
the more general stationary phase conditions of Eq.(\ref{grad}) for
selecting the effective oscillation periods of the coupling. Due to the
equivalence between $q_x$ and $q_y$, it is clear from figures \ref{FSN3} and
\ref{FSN6} that all wave vectors $q_\perp$ satisfying the stationary
phase conditions are located at the origin of the 2-dimensional
Brillouin zone, {\it i.e.}, $q_x = q_y = 0$. Therefore, the oscillation
period of the propagator $g_{0 \, N_c}$ is associated with the value of
$q_\perp$ determined by the solution of Eq.(\ref{cosq}) for $q_x = q_y =
0$. 

\begin{figure}
\begin{center}\mbox{\psboxscaled{800}{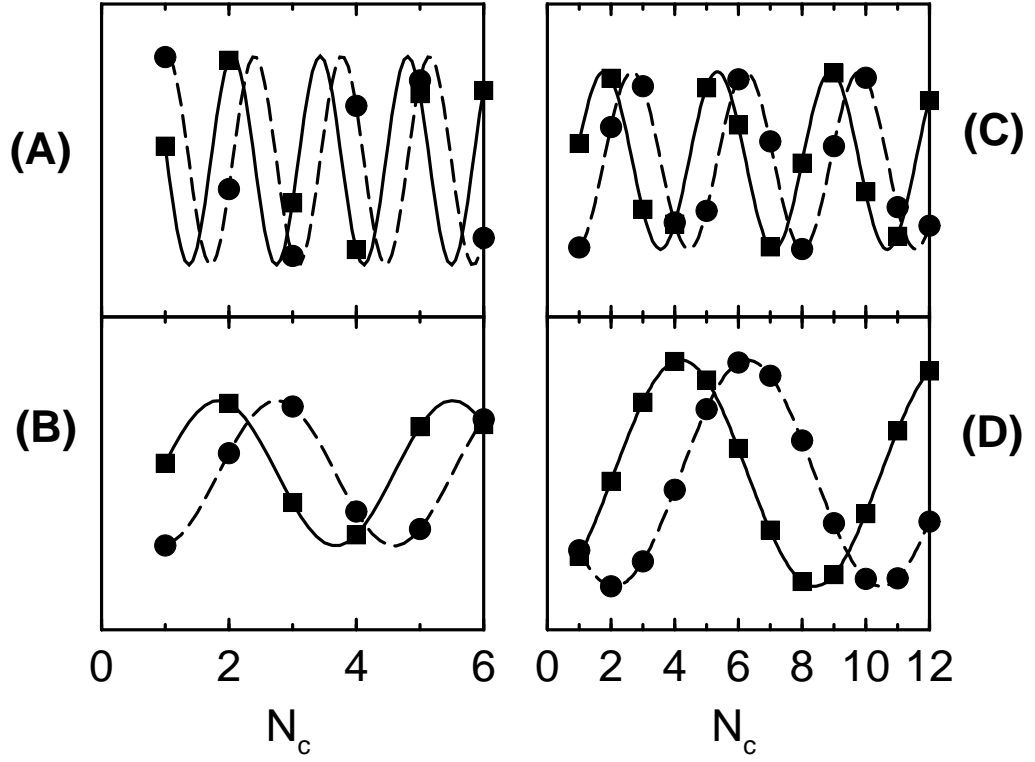}}\end{center}
\caption{$g^{\alpha \, \alpha^\prime}_{N_c \, 0} \,\, g^{\alpha^\prime \, 
\alpha}_{0 \, N_c}$
(in arbitrary units) as a function of the number of cells $N_c$ for 
some intracell parameters. $\alpha = \alpha^\prime = 1$. 
The full lines with squares are the real part of the
functions whereas the dashed lines with circles are the imaginary
part. The symbols represent the integer values of $N_c$, whereas the 
lines are a guide to the eye. (A) $N_A = 1$ and $N_B = 2$; (B) $N_A = 2$
and $N_B = 1$; (C) $N_A = 3$ and $N_B = 3$; (D) $N_A = 4$ and $N_B =
2$. In cases (A) and (B) the Fermi level is $E_F = -1.6$ whereas $E_F =
-1.35$ in cases (C) and (D).}  
\label{oscillations}
\end{figure}

According to Eqs. (\ref{j}) and (\ref{Gg}), the coupling is a
function of the propagators $g$ between the magnetic planes. In the case
considered here, $g^{\alpha \, \alpha^\prime}_{0 \, N_c}$ and
$g^{\alpha^\prime \, \alpha}_{N_c \, 0}$ are the propagators
involved. We recall that in this case the coupling depends on the
product $g^{\alpha \, \alpha^\prime}_{0 \, N_c} \, g^{\alpha^\prime \,
\alpha}_{N_c \, 0}$ and bearing in mind that both propagators oscillate
with the same periods, the coupling oscillates with an effective wave
vector which is twice that for $g_{0 \, N_c}$. The corresponding
periods of the coupling are then given by 
$\Lambda(N_A,N_B) = \pi / q_\perp$. The periods depend on the intra-cell
composition determined by $N_A$ and $N_B$ as can be seen by the change
in caliper of the extremal wave vectors spanning the Fermi surface in
the centre of the Brillouin zones. Following Eq.(\ref{cosq}), the
periods are displayed in Table \ref{tabelperiods} and the product
$g^{\alpha \, \alpha^\prime}_{N_c \, 0} \,\, g^{\alpha^\prime \,
\alpha}_{0 \, N _c}$ as a function of $N_c$ for different
intracell parameters and for $\alpha = \alpha^\prime = 1$ is shown in 
figure \ref{oscillations}. Note the agreement between the periods in 
figure \ref{oscillations} and the ones in Table \ref{tabelperiods}. 

\section{discussion and conclusions}

The results presented above indicate that the oscillation periods of the 
interlayer coupling across superlatticed spacers are related to the shape 
of the Fermi surface when the coupling is investigated as a function of 
the number of cells of the superlattice. Therefore, a change in the 
composition of the superlattice enables one to deform the corresponding 
Fermi surface and consequently tune the oscillation periods of the coupling. 
Moreover, although the present analysis is not concerned with the
strength of the oscillation periods, it should be mentioned that the
change in intra-cell parameters not only alters the relevant wave
vectors but also the respective curvatures of the Fermi
surface. Bearing in mind the dependence of the oscillation amplitude on
such curvatures, the contribution of periods may be enhanced by the
variation of the parameters $N_A$ and $N_B$. Non-RKKY periods, for
instance, which in some cases contribute weakly to the oscillations
\cite{non-rkky,fundamental}, could become more relevant through the
variation of those parameters. 

It is hoped that the simple case treated here reflects what occurs in 
real systems, which may be confirmed by future experiments. Alloys that, 
under certain critical temperatures, present ordered phases may be used to
test the veracity of the present claims. In the ordered phase, Cu-Zn 
alloys, for instance, have a BCC structure consisting of interpenetrating 
simple cubic sublattices \cite{kittel}. In this case, cells formed by one 
plane of Cu and one plane of Zn are periodically repeated, corresponding 
to a possible realization of the system proposed in this communication. 
The lattice parameters of the ordered Cu-Zn alloy and that of Fe are 
similar, which may allow them to be grown in layered structures. Although
the oscillation periods of the coupling cannot be tuned in this system, it 
is suggested that the periodicities associated with the corresponding Fermi 
surface will become evident only when plotted as a function of even 
(or odd) number of spacer layers. 
Alternatively, by growing superlattices of non-magnetic metals, {\it e.g.} 
noble metals, in wedge-shaped systems \cite{israel,aliev} one can change 
the intracell composition of the system to test whether the periods of the 
coupling can be tuned, as the present calculations indicate.

\section*{Acknowledgements}

I wish to acknowledge Professor D. M. Edwards, Dr. R. B. Muniz, Dr. J. 
d'Albuquerque e Castro and Dr. R. Wright for their invaluable contributions 
to this work. Also, I am grateful to CNPq of Brazil for financial 
support.
\newpage
\newpage
\newpage
\newpage
\newpage
\newpage

\begin{table}
\begin{tabular}{cc}
$N = 3$ & $N = 6$                              \\ \hline
$\Lambda(0,3) = 1.13$  & $\Lambda(2,4) = 1.69$ \\ 
$\Lambda(1,2) = 1.37$  & $\Lambda(3,3) = 3.56$ \\ 
$\Lambda(2,1) = 3.67$  & $\Lambda(4,2) = 8.39$ \\ 
$\Lambda(3,0) = 9.33$  &                       \\ 
\end{tabular}
\caption{{\it Oscillation periods $\Lambda (N_A,N_B)$ (in units of $D$)
of the coupling as a function of the number of cells $N_c$ for
different values of $N_A$ and $N_B$. }}
\label{tabelperiods}
\end{table}

\newpage

\end{document}